\documentclass[journal]{IEEEtran}
\usepackage{cite}

\ifCLASSINFOpdf
  \usepackage[pdftex]{graphicx}
\else
  \usepackage[dvips]{graphicx}
\fi

\usepackage[cmex10]{amsmath}
\usepackage{amsfonts}
\usepackage{amssymb}
\usepackage{graphicx}
\usepackage{subfigure}
\usepackage{color,soul}

\usepackage{algorithmic}

\usepackage{verbatim}
\usepackage{dcon}

\makeatletter

\newcommand{\Rmnum}[1]{\expandafter\@slowromancap\romannumeral #1@}
\makeatother
\usepackage{setspace}
\begin{document}
%
\title{Active Consensus over Sensor Networks \\ via Randomized Communication}
%

\author{Lei~Chen, Jeff~Frolik 
}

%
%

\maketitle

\begin{abstract}

Distributed consensus has been widely studied for sensor network applications.
Whereas the asymptotic convergence rate
has been extensively explored in prior work, other important and
practical issues, including energy efficiency and link reliability, have received relatively
little attention. In this paper, we present a
distributed consensus approach that can achieve a good balance
between convergence rate and energy efficiency. 
 
The approach selects a subset of links
that significantly contribute to the formation of consensus at each
iteration, thus adapting the network's topology dynamically to the
changes of the sensor states.
A global optimization problem is formulated for optimal link selection, which is subsequently factorized into sub-problems that can be solved locally, and practically via approximation. An algorithm is derived to solve the approximation efficiently, using quadratic programming (QP) relaxation and random sampling.
Simulations on networks of different types demonstrate that the proposed method reduces the communication energy costs without significantly impacting the convergence rate and that the approach is robust to link failures.
\end{abstract}



\IEEEpeerreviewmaketitle

\section{Introduction}
\label{sec:intro}

\IEEEPARstart{T}{he} past decade has witnessed the growing use of wireless sensor networks
(WSNs) in both industrial and scientific applications, including
environment monitoring, structural monitoring, and 
emergence detection.
With these networks in place, there is increasing interest in developing fast, robust, and energy efficient means to collect and aggregate data for both decision and control purposes. Collectively this area of research is called consensus forming or sensor fusion and can be roughly categorized as being either centralized or distributed among the sensor nodes. Distributed consensus, the focus of the work herein, has the advantages of improved robustness, scalability, and energy efficiency.  

Much of the work to date in distributed consensus has been focused on algorithms and their convergence behavior. 
Xiao and Boyd~\cite{Lin2003} considered distributed consensus through linear iterations. 
Boyd~\cite{Boyd2006} extended the work to randomized gossip. 
In both approaches, the convergence rate was shown to be closely related to the spectral structure of the network, i.e., the eigen-system of the network topology.
Franceschelli \etal~\cite{France} proposed a new decentralized gossip algorithm based on broadcasts.
Salehi and Jadbabaie~\cite{Salehi2007} studied the asymptotic behavior of
the consensus algorithms over switching graph.
Recently, Chen \etal~\cite{Chen1,Chen2} studied a new formulation for distributed consensus based on state derivative, they also proposed an efficient approach which allows each node in a network to individually determine whether consensus has been attained.
Pereira~\cite{Pereira2011} evaluated the convergence of the consensus
algorithm over networks with random asymmetric topology. 
Hatime~\cite{Hatime} investigated the impact of topological characteristics on consensus building in multiagent systems. 
%
%

Guided by these convergence analyses, various methods have been
developed to design the optimal network structure that can achieve
fastest convergence. Xiao and Boyd~\cite{Lin2003,Boyd2003} developed
an optimization algorithm that minimizes the second largest eigenvalue.
Das and Mesbahi~\cite{Das2006} described a linear estimation algorithm
based on distributed consensus by exploiting the clustering
structure.
Jin and Murray~\cite{Jin2007} proposed a new method that introduces virtual links between skeleton nodes to speed up convergence.
Jakovetic \etal~\cite{Jak2010} presented a new formulation that addresses correlated random topologies.
Jafarizadeh \etal~\cite{Jafari} presented an analytical solution for the problem of fastest distributed consensus for a sensor network that is composed by two different symmetric star sensor networks.
The previous work mentioned above focuses on theoretical analysis, where important practical aspects (e.g., energy efficiency) are often neglected. In this work, we focus on improving the energy efficiency of the network communication, i.e., reducing the energy cost while maintaining a reasonable convergence performance, which is crucial for the efficacy of a practical sensor network.

Earlier work on distributed consensus~\cite{Lin2003,Boyd2003,Das2006,Jin2007} also focused on a static networks, optimizing the updating weights solely based on the network topology. Recently, randomized algorithms~\cite{Boyd2006,Kar2009,Pereira2011} that
can generate graphs with varying topologies have become increasingly
popular. However, in such an algorithm, the dynamic changes of the network topology typically follows a stationary process, which is devised without utilizing run-time information, that is, the information used is only from the initial network design.

In this paper, we take a different approach, exploiting the
information available in run-time, i.e., the states maintained by the
sensor nodes, to adapt the network topology on the fly. 
We observed that the usefulness of the various network links for obtaining consensus largely
depends on run-time status. 
For example, the communication between two nodes with disparate states 
is more useful than that between nodes with similar states.
Motivated by this intuition, we formulate an energy-constrained optimization problem,
which seeks the most useful subset of links for consensus forming at each iteration, under a constraint on energy costs.
In addition, we derive a randomized algorithm that can
approximately solve this global problem in a distributed manner, by
decomposing it into locally solvable problems.
The active utilization of run-time states to optimize the
communication topology along with the consensus building process
clearly distinguishes this work from other schemes, and we thus call
it \emph{active consensus}.
Furthermore, as a practical issue, the robustness to link failures are also studied. This work extends earlier developments along these lines~\cite{Chen_secon} by considering new topologies and investigating the robustness of the approach to link failures.

It is worth reiterating that the primary goal of this work is to address the practical problem of increasing energy efficiency. To test the efficacy of the proposed scheme, we performed simulation under a variety of conditions. The experimental results demonstrated that this method can substantially reduce the overall energy consumption, without significantly degrading the convergence time.

The remainder of this paper is organized as follows. 
In Section~\ref{sec:theory}, we revisit the theoretical analysis of
the consensus algorithm and reveal its relations to a distributed
optimization problem. 
In Section~\ref{sec:comm}, we formulate an optimization approach that selects the appropriate communication links in run-time, and then develop a randomized approximation
algorithm that can be implemented locally.
In Section~\ref{sec:simulation}, we present the simulation results
under different topologies and channel conditions, which demonstrate the energy efficiency improvement achieved by the proposed method.
Finally, we conclude the work in Section~\ref{sec:conclusion}.


\section{Theory}
\label{sec:theory}

In this section, we first formalize a communication model 
for a sensor network in Section~\ref{sub:comm_model}, and thereon
analyze the process of distributed average consensus in Section~\ref{sub:dac}.
Then, we discuss a different perspective in
Section~\ref{sub:optimview}, which relates the formation of consensus
to a distributed optimization procedure. 
As we shall see in Section~\ref{sec:comm}, this relation provides theoretical
insight to guide the design of an energy-efficient communication
scheme.

\subsection{Iterative Linear Updates}
\label{sub:comm_model}

Consider a sensor network comprised of a set of nodes, denoted by $V$,
and a set of symmetric links\footnote{A network is symmetric if
  each link of the network is bidirectional, meaning that either end
  of the link can receive information of the other.} between them,
denoted by $E$.
Each node maintains a state value, which is
dynamically updated based on the information from neighboring nodes. 
In this work, we assume time is discretized into time steps, and all the nodes of the network talk to their neighbors simultaneously at each time step (i.e., synchronized communication).
In addition, we focus on the peer-to-peer communications that are typical
in distributed sensor networks, where each communication step is between two
specific nodes.
Let $x_v(t)$ denote the state value maintained by the node $v$ at time
$t$, which can be intuitively understood as the node's estimate of the consensus value. For each node $v \in V$, $x_v(0)$ is initialized to
be the value measured by the corresponding sensor, which is then
iteratively updated using a linear combination of the values received
from the neighbors, as follows
\begin{equation} \label{eq:update0}
    x_v(t) = x_v(t-1) - 
    \delta \sum_{u \in \nset_v} (x_v(t-1) - x_u(t-1)). 
\end{equation}
Here, $\nset_v = \{u : \{v, u\} \in E\}$ is the set of neighbors of
$v$, and $\delta$ is the step size. 
As we shall see later, the asymptotic behavior of this updating
process is largely determined by the \emph{Laplacian matrix} $\mL$, 
which is defined as follows
\begin{equation}
    \mL(u, v) = \begin{cases}
        d_v & (u = v), \\
        -1 & (\{u, v\} \in E), \\
        0 & (\text{otherwise}).
    \end{cases} 
\end{equation}
Here, $d_v = |\nset(v)|$ is the degree of $v$ (i.e., the number of neighbors of $v$). With the Laplacian matrix $\mL$, the updating formula in 
Eq. (\ref{eq:update0}) can be written in a vector form as
\begin{equation} \label{eq:update1}
    \vx(t) = (\mI - \delta \mL) \vx(t-1).
\end{equation}
Here, $\vx(t)$ is a $n$-dimensional vector composed of the states of
all nodes at time $t$. For conciseness, we use 
$\mW(\mL, \delta) \triangleq \mI - \delta \mL$ to denote the update matrix, which has
\begin{equation} \label{eq:Weigen}
    \mW(\mL, \delta) \ve_i = 
    (\mI - \delta \mL) \ve_i =
    (1 - \delta \lambda_i(\mL)) \ve_i.
\end{equation}
This implies that $\mW(\mL, \delta)$ has eigenvalues
$1 - \delta \lambda_1(\mL), \ldots, 1 - \delta \lambda_n(\mL)$, which
are associated with the same eigenvectors as $\mL$, namely 
$\ve_1, \ldots, \ve_n$. In particular, since $\lambda_1(\mL) = 0$
and $\ve_1 = \vone$, we have $\mW \vone = \vone$.

\subsection{Distributed Average Consensus}
\label{sub:dac}

Next, we consider how the updating process evolves over time and show that under certain conditions, the states of all nodes will
reach an average value, called the \emph{consensus}. We note that this
is the result that underpins the distributed average consensus
methdology, which we review here so as to provide a theoretical
foundation for further discussion.

Let $S_1$ denote the subspace spanned by $\ve_1 = \vone$, and $D
\triangleq S_1^\perp$ denote its orthogonal complement, which is
spanned by the remaining eigenvectors $\ve_2, \ldots, \ve_n$.
Clearly, each vector $\vx \in D$ satisfies $\vone^T \vx = 0$.
Then each state vector $\vx(t)$ can be uniquely decomposed into a
linear combination of two components: one in $S_1$ and the other in
$D$, as follows
\begin{equation} \label{eq:decomp0}
    \vx(t) = \mu(t) \vone + P_D \vx(t).
\end{equation}
Here, $\mu(t)$ is the average of the values in $\vx(t)$, and $P_D
\vx(t)$ the projection of $\vx(t)$ onto $D$.
The component $\mu(t) \vone$ has the same value at all nodes, which we
call the consensus component; while the other component $P_D
\vx(t)$ reflects the differences between nodes, which we call the difference component. 
Then at time $t + 1$, the updated state vector becomes
\begin{equation}
    \vx(t+1) = \mW(\mL, \delta) \vx(t) = \mu(t) \vone + \mW(\mL,
    \delta) P_D \vx(t).     
\end{equation}
Here, we utilize the fact that $\mW(\mL, \delta) \vone = \vone$.
It can be easily verified that this constitutes an orthogonal
decomposition of $\vx(t+1)$ along $S_1$ and $D$. On the other hand,
such decomposition of $\vx(t+1)$ can be expressed as
\begin{equation}
    \vx(t+1) = \mu(t+1) \vone + P_D \vx(t+1).    
\end{equation}
Therefore, we have
\begin{align}
    \mu(t+1) &= \mu(t), \\
    P_D \vx(t+1) &= \mW(\mL, \delta) P_D \vx(t).
\end{align}
This implies that the consensus component is fixed over
time. Therefore, we can use $\mu \triangleq \mu(t)$
to indicate this fixed average value. 
Applying the formulas above recursively results in
\begin{equation}
    \vx(t) = \mu \vone + \mW^t(\mL, \delta) P_D \vx(0).    
\end{equation}
Here, the evolution of the difference component, 
\ie,~$P_D \vx(t)$, depends on the spectrum 
of $\mW(\mL, \delta)$. Specifically, we have
\begin{align}
    \| P_D \vx(t) \| 
    &= \| \mW^t(\mL, \delta) P_D \vx(0) \| \notag\\
    &\le \| \mW(\mL, \delta) \|_D^t
    \| P_D \vx(0) \|.
\end{align}
Here, $\| \mW \|_D$ indicates the operator norm of
$\mW$ \wrt~the subspace $D$, \ie,~the maximum magnitude of its
eigenvalues. According to Eq. (\ref{eq:Weigen}), this norm is related
to the eigenvalues of $\mL$, as
\begin{equation}
    \| \mW(\mL, \delta) \|_D = 
    \max \{ |1 - \delta \lambda_2(\mL)|, 
    |1 - \delta \lambda_n(\mL)| \}.
\end{equation}
Clearly, when $\| \mW(\mL, \delta) \| < 1$, $P_D \vx(t)$ attenuates to
$\vzero$ as $t \ato \infty$. Consequently, 
$\vx(t)$ approaches $\mu \vone$, the consensus status, in which all
nodes have the same state $\mu$. We summarize the analysis above in
the following theorem.
\begin{theorem} \label{thm:consensus}
    The updating process described in Eq. (\ref{eq:update1}) converges
    in norm to $\mu \vone$ when $\| \mW(\mL, \delta) \|_D < 1$, or 
    equivalently, when both $|1 - \delta \lambda_2(\mL)|$ and 
    $|1 - \delta \lambda_n(\mL)|$ are less than $1$.
    Here, $\mu$ is the average of the values in $\vx(0)$.
\end{theorem}
Furthermore, the convergence time (up to some precision) is
proportional to the value of 
$1 / \log \left( \| \mW(\mL,\delta)\|_D^{-1} \right)$.
Therefore, one can attain the fastest convergence by minimizing 
$\| \mW(\mL, \delta) \|_D$. 
Particularly, given $\mL$, the optimal step size is 
\begin{equation} \label{eq:optstep}
    \hat\delta = \frac{2}{\lambda_2(\mL) + \lambda_n(\mL)}.
\end{equation}
We note that Boyd \etal~\cite{Boyd2003} derived a similar result on
optimal choice of $\delta$.

\subsection{An Optimization-Based Perspective}
\label{sub:optimview}

Given a state vector $\vx$, the sum of squared
differences between neighboring nodes can be expressed concisely with
a Laplacian matrix, as
\begin{equation}
    \frac{1}{2} \sum_v \sum_{u \in \nset_v} (x_v - x_u)^2 
    = \vx^T \mL \vx.
\end{equation}
Suppose the condition given by Theorem~\ref{thm:consensus} is
satisfied, it is a natural intuition that the differences between
neighbor states gradually diminish as the updating proceeds. 
The following theorem establishes this intuition rigorously as a
fact, stating that the value of $\vx(t)^T \mL \vx(t)$ decreases 
as $t$ increases. 
\begin{theorem}
    Let $\vx(t)$ be an updating process described in
    Eq.~(\ref{eq:update1}), then
    \begin{equation} \label{eq:Lnorm2_shr}
        \vx(t+1) \mL \vx(t+1) \le
        \| \mW(\mL, \delta) \|_D^2 \cdot (\vx(t) \mL \vx(t)).
    \end{equation}
    In particular, if the condition in Theorem~\ref{thm:consensus} is
    satisfied, namely $\|\mW(\mL, \delta)\|_D < 1$, then
    $\vx(t) \mL \vx(t)$ is a decreasing sequence, provided that 
    the difference component of $\vx(0)$ is non-zero.
\end{theorem}
\vspace{5pt}

This theorem together with the fact that $\mu(t) \equiv \mu$ suggests
that the updating process described in Eq.~(\ref{eq:update1}) is
  actually minimizing the objective function $f_\mL(\vx) = \vx^T \mL
  \vx$ in a distributed fashion, subject to the constraint $(\vone^T
  \vx) / n = \mu$.
This optimization problem has a unique optima $\hat{\vx} = \mu \vone$.
Moreover, Eq.~(\ref{eq:Lnorm2_shr}) provides the convergence rate of
the objective value.


\section{Energy-Constrained Communication}
\label{sec:comm}

\begin{figure}[t]
    \centering
    \includegraphics[width=0.45\textwidth]{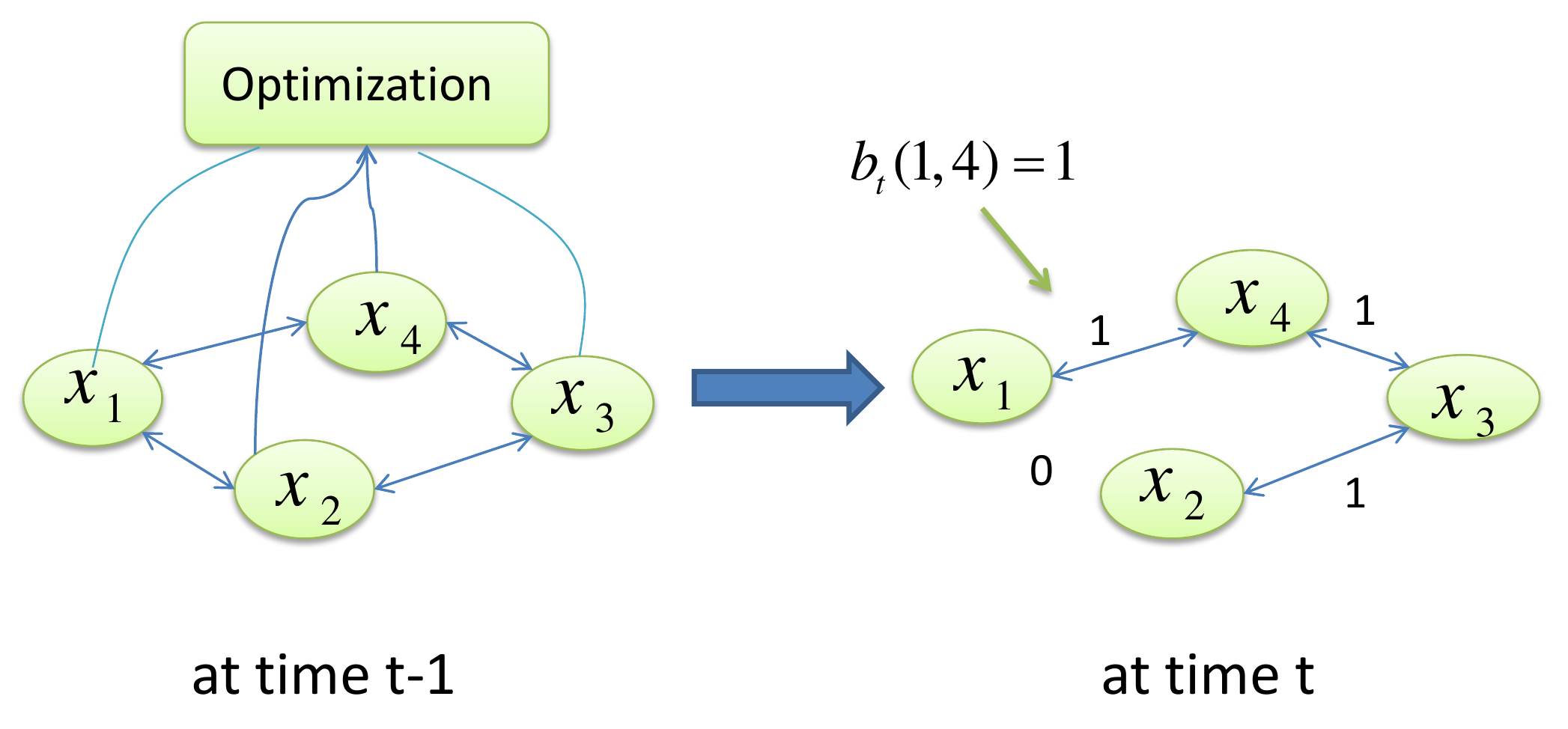}
    \caption{This figure illustrates the proposed global optimization
      scheme on a network comprised of four nodes. At each iteration,
      previous states of the nodes are utilized as the input to the
      optimization problem given in Eq.(\ref{eq:key_global}), which finds
      the optimal subset of links that can decrease the differences
      between neighbors to the greatest extend, subject to the energy
      constraint. In this case, the optimal solution chooses three
      links: $\{a,b\}, \{b, c\}$, and $\{c,d\}$. They are used in the next
      iteration of updating.}
    \label{fig:global_op}
\end{figure}

Wireless sensor networks, especially those deployed on field, typically battery powered or utilize energy harvesting, and thus are
usually subject to energy constraints. The primary goal of this work
is to derive a practical communication scheme to improve the energy
efficiency of distributed consensus. Inspired by the theoretical analysis in previous section,
we present two approaches in this section. The basic idea
underlying both methods is to adaptively choose a subset of links at
each iteration, such that the information exchanged via these links
contributes most to the formation of consensus.  
Specifically, in Section~\ref{sec:globalopt}, we first formalize the
selection of links as an optimization problem and develop a methods that
chooses the optimal subset of links by solving this problem.
To make the algorithm practical, in Section~\ref{sec:localopt},
we then factorize the optimization problem into sub-problems that
can be solved locally using a randomized method.

\subsection{Globally Optimal Link Selection}
\label{sec:globalopt}

As discussed in Section~\ref{sub:optimview}, building consensus over
a network can be considered as minimizing the metric 
$\vx(t)^T \mL \vx(t)$. 
In this sense, we can measure the effectiveness of an updating
iteration in terms of how much it decreases this metric. 
Suppose we only use a subset of links at time $t$ to perform the
update, then the updating formula can be written as
\begin{equation} \label{eq:bupdate0}
    x_v(t) = x_v(t-1) - \delta 
    \sum_{u \in \nset_v} b_t(\{v,u\}) (x_v(t-1) - x_u(t-1)).
\end{equation}
Here, $b_t(\{v,u\})$ indicates whether the link $\{v, u\}$ is used in
the communication at time $t$, \ie~$b_t(\{v, u\}) = 1$ if this link is
used and $b_t(\{v,u\}) = 0$ otherwise.
Hence, $\vx(t)^T \mL \vx(t)$ depends on the choice of links, which can
be captured by $\vb_t = b_t(e)$, where $e = \{u, v\} \in E$ the vector comprised of all the link indicators. Following this argument, we can choose the
most effective subset of links, by finding the optimal $\vb_t$ in
terms of minimizing $\vx(t)^T \mL \vx(t)$.
To derive a vector form of this objective function, 
we let $\mU_{\vx(t)}$ be an $|V| \times |E|$ matrix given by 
\begin{equation}
    \mU_{\vx(t)}(v, e) = \begin{cases}
        x_v(t) - x_u(t) & (\{v, u\} \in E) \\
        0 & (\text{otherwise}).
    \end{cases}
\end{equation}
Then, Eq. (\ref{eq:bupdate0}) can be written into
\begin{equation}
    \vx(t) = \vx(t-1) - \delta \mU_{\vx(t-1)} \vb_t.
\end{equation}
Hence, given $\vx(t-1)$, the objective function of choosing the
optimal subset of links can be written as
\begin{align} \label{eq:key_global}
    Q(\vb_t) 
    &= \frac{1}{2} \sum_{v \in V} \sum_{u \in \nset_v}
    (x_v(t) - x_u(t))^2 = \vx(t)^T \mL \vx(t) \notag \\
    &= (\vx(t-1) - \delta \mU_{\vx(t-1)} \vb_t)^T \mL
    (\vx(t-1) - \delta \mU_{\vx(t-1)} \vb_t)  \notag \\
    &= \delta^2 \vb_t \mH_{\vx(t-1)} \vb_t
    - 2\delta \vf_{\vx(t-1)}^T \vb_t + \mathrm{const.} 
\end{align}
Here, we introduce $\mH_{\vx(t-1)}$ and $\vf_{\vx(t-1)}$ to simplify
the notation, which are defined to be
\begin{align*}
    \mH_{\vx(t-1)} &= \mU_{\vx(t-1)}^T \mL \mU_{\vx(t-1)}, \\
    \vf_{\vx(t-1)} &= \mU_{\vx(t-1)}^T \mL \vx(t-1).
\end{align*}

In general, using more links in the communication tends to speed up
the convergence. As a result, purely pursuing fastest convergence
would lead to the selection of many links, incurring high energy
consumption. In practice, it is usually more desirable to seek a
balance between convergence rate the energy cost. To this end, we
impose a cost constraint to this optimization problem. Particularly,
we associate each link $e \in E$ with a cost value $c_e$, which
reflects the energy needed to communicate via this link. 
Let $\vc = (c_e)_{e \in E}$ be the vector composed of the cost values
of all links. Then, the constraint can be expressed as 
\begin{equation} \label{eq:global_cc}
    \vc^T \vb_t = \sum_{e \in E} c_e b_t(e) \le C.    
\end{equation}
This means that the total communication cost at time $t$ should not
exceed $C$. Combining Eq. (\ref{eq:key_global}) and
Eq. (\ref{eq:global_cc}), we obtain a constrained optimization problem, 
given by
\begin{align}
    \text{minimize } & 
    \frac{1}{2} \delta^2 \vb_t \mH_{\vx(t-1)} \vb_t
    - \delta \vf_{\vx(t-1)}^T \vb_t, \\   
    \text{s.t. } &
    \vc^T \vb_t \le C.
\end{align}
Note that each entry of $\vb_t$ can only take a value from $\{0, 1\}$.
Hence, this is a binary integer programming 
problem\footnote{A binary integer programming problem is an
  optimization problem, of which each variable can only take a binary
  value, \ie~either $0$ or $1$.}.
In general, finding the optimal solution to this problem is
NP-hard. 
However, taking advantage of the quadratic form of the problem, we
derive an efficient algorithm as follows. 
First, through a relaxation that allows the value of $b_t(e)$ to be
any real number in $[0, 1]$, this problem reduces to a quadratic
programming (QP) problem with linear constraints that can be readily
solved. Denote the solution to the QP as $\vp_t$. 
To turn this into a binary solution $\vb_t$, we treat each value of
$p_t(e)$ as a probability, and thus obtain the value of $b_t(e)$ by
random sampling. Particularly, for each $e \in E$, we draw $b_t(e) \in
\{0, 1\}$ with $P(b_t(e) = 1) = p_t(e)$. For instance, if $p_t(e) =
0.7$, then we turn on the edge $e$ at time $t$ with a chance of $70\%$. An illustration of the proposed global optimization scheme is shown in Fig~\ref{fig:global_op}.

We note that $\vb_t$ obtained using this approximate method is not
necessarily the optimal solution to the original binary integer
programming problem, and one can further refine this solution using
MCMC~\cite{mcmc} simulated annealing. 
However, we find that this is not necessary, as the solution $\vb_t$
derived via sampling is very close to the true optima in most cases. Suppose $\vb_t^*$ is the true optimal solution to the original problem, \ie~it attains the minimum objective among all binary vectors, then $Q(\vb_t^*) \le
Q(\vb_t)$. Moreover, $\vp_t$ is the optimal solution to the relaxed
problem, \ie~it attains the minimum within a larger domain that
allows real values, implying that $Q(\vp_t) \le Q(\vb_t^*)$. Together,
we have
\begin{equation}
    Q(\vb_t) \ge Q(\vb_t^*) \ge Q(\vp_t).    
\end{equation}
In experiments, we found that $Q(\vb_t)$ is very close to $Q(\vp_t)$ in
most cases, implying that $Q(\vb_t)$ is even more close to
$Q(\vb_t^*)$, as $Q(\vp_t)$ provides a lower bound to the minimum of
the original problem.
This, in other words, means that $Q(\vb_t)$ is close to the true optimum.

\subsection{Locally Optimal Link Selection}
\label{sec:localopt}

As we shall see in next section, the algorithm described above is very
effective in optimizing the network topology under cost constraints. 
However, the reliance on solving a QP problem over the joint state
vector makes it impractical in a distributed context. To address this
issue, we consider an approximation, with which, the
optimization problem can be decomposed into a collection of 
sub-problems that can be solved locally by nodes.

Given $\vx(t-1)$, we can rewrite the objective function $Q(\vb_t)$
in Eq. (\ref{eq:key_global}) as follows.
\begin{equation}
    Q(\vb_t) = \sum_{v \in V} Q_v(\vb_t),
\end{equation}
with
\begin{equation}
    Q_v(\vb_t) = \frac{1}{2} \sum_{u \in \nset_v} ( x_v(t) - x_u(t))^2.         
\end{equation}
Here, $x_v(t)$, as given by Eq. (\ref{eq:bupdate0}), is the value of
node $v$ at next time step $t$, and therefore $Q_v(\vb_t)$ is the sum
of the squared differences between the value of $v$ and those of its
neighbors.

In this way, we decompose $Q(\vb_t)$, the overall objective, into
the sum of local objectives $Q_v(\vb_t)$, each associated with a node
$v$. Our goal here is to divide the original optimization problem into
ones that can be solved separately and locally by the sensor nodes, \ie~each node can
solve its corresponding problem solely based on the information that
it has.

However, $Q_v(\vb_t)$ cannot be optimized locally with its current
form. Note that $Q_v(\vb_t)$ depends on $x_u(t)$ for each $u \in
\nset_v$, which in itself depends on its neighbor values, as
\begin{equation*}
    x_u(t) = x_u(t-1) - \delta
    \sum_{w \in \nset_u} b_t(\{u,w\}) (x_u(t-1) - x_w(t-1)).
\end{equation*}
The issue here is that $w$ is not necessarily $v$'s neighbor, and for
such nodes, node $v$ has no idea of their values. To address this
issue, we simply ignore non-neigbors of $v$, and derive an
approximated updating formula below
\begin{equation}
    \tilde{x}_{v:u}(t) = x_u(t-1) - \delta\sum_{w \in \nset_u \cap \nset_v} (x_w(t-1) - x_u(t-1)).        
\end{equation}
Here, $\nset_u \cap \nset_v$ is the set of neighbors shared by both
$u$ and $v$. The node $v$ uses the values of these nodes to make an
approximate prediction of what the value $u$ is at time $t$, denoted
by $\tilde{x}_{v:u}(t)$, resulting in the following approximate local
objective:
\begin{equation} \label{eq:localQv}
    \tilde{Q}_v(\vb_t(E_v)) = \frac{1}{2}
    \sum_{u \in \nset_v} (x_v(t) - \tilde{x}_{v:u}(t))^2.
\end{equation}
Here, $E_v = \{(v, u): u \in \nset_v\}$ is the subset of edges
incident with node $v$, and $\vb_t(E_v)$ is the corresponding
sub-vector $\vb_t$, which consists of all the link indicators for $v$. 

We also break the original total cost constraint
$\vc^T \vb_t \le C$ into local constraints as follows,
so as to make the sub-problems locally solvable.
\begin{equation} \label{eq:lc}
    \sum_{e \in E_v} c_e b_t(e) \le C_v.
\end{equation}
Here, $C_v$ is the cost bound individually imposed on node $v$,
which is proportional to the degree of node $v$.

We have obtained a local optimization problem for each node
$v$, \ie~to minimize $\tilde{Q}_v$ given by Eq. (\ref{eq:localQv}),
subject to the local constraint given by Eq. (\ref{eq:lc}). This
problem can be solved separately by the node $v$, without consulting
other nodes.
This is an integer programming problem, which is difficult to
solve directly.  Again, we relax the binary value $b_t(e) \in \{0,
1\}$ to a real value $p_t(e) \in [0, 1]$, thus turning it into a
QP problem. Here, $p_t(e)$ can be considered as the probability that 
the link $e$ should be used at time $t$. 
Note that for each edge $e = \{u, v\} \in E$, we obtain
two probability values for $e$, respectively when optimizing $Q_u$ and
$Q_v$. 
These values are not the same in general. We take a simple method, 
using the average of them to be the probability of setting $b_t(e) =
1$.


\section{Demonstration of Approach}
\label{sec:simulation}

\begin{figure*}
    \centering
    \subfigure[Topology of an example network]
    {\includegraphics[width=0.31\textwidth]{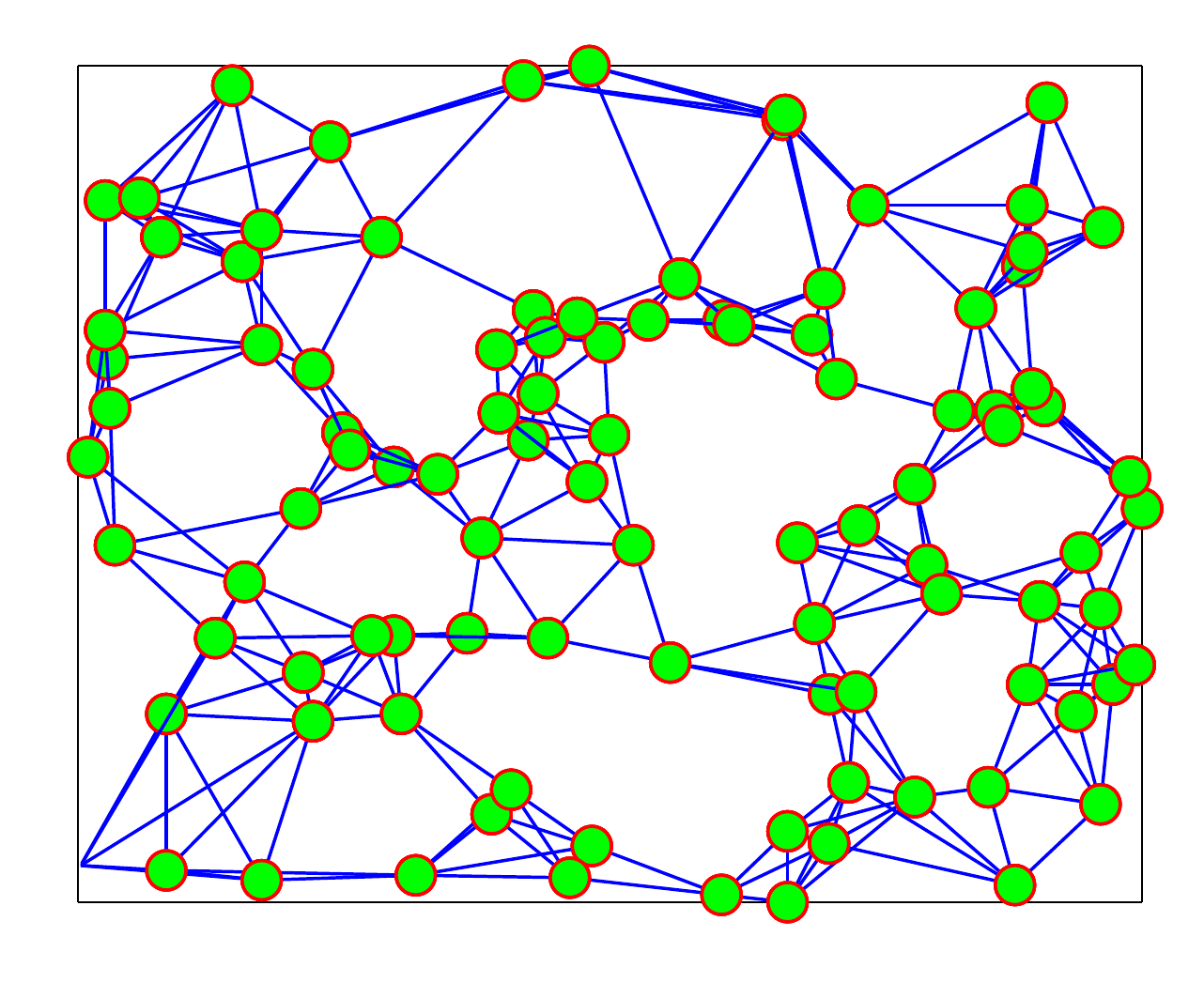}}
    \subfigure[Cost ratio as a function of $\alpha$]
    {\includegraphics[width=0.31\textwidth]{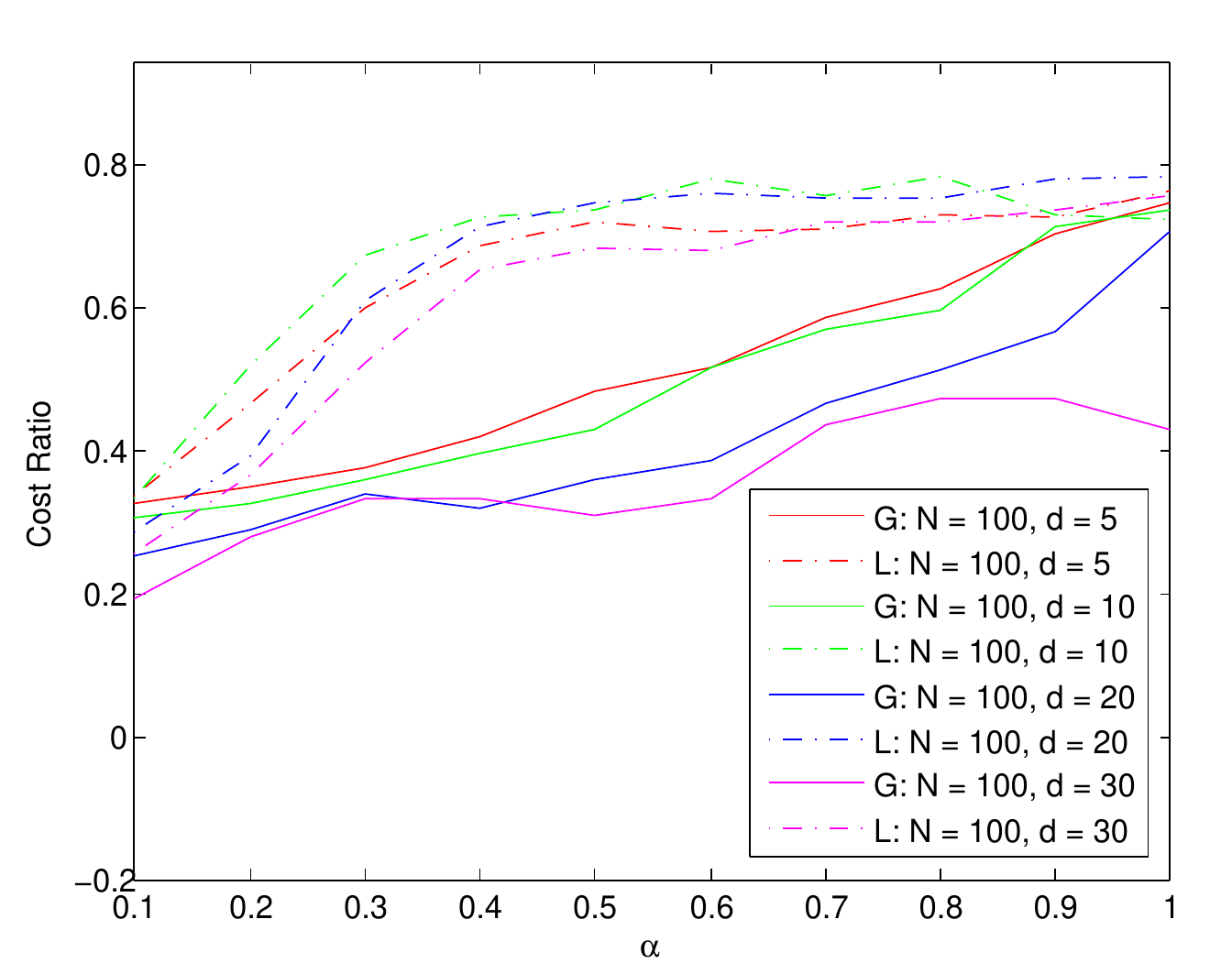}}
    \subfigure[Time ratio as a function of $\alpha$]
    {\includegraphics[width=0.325\textwidth]{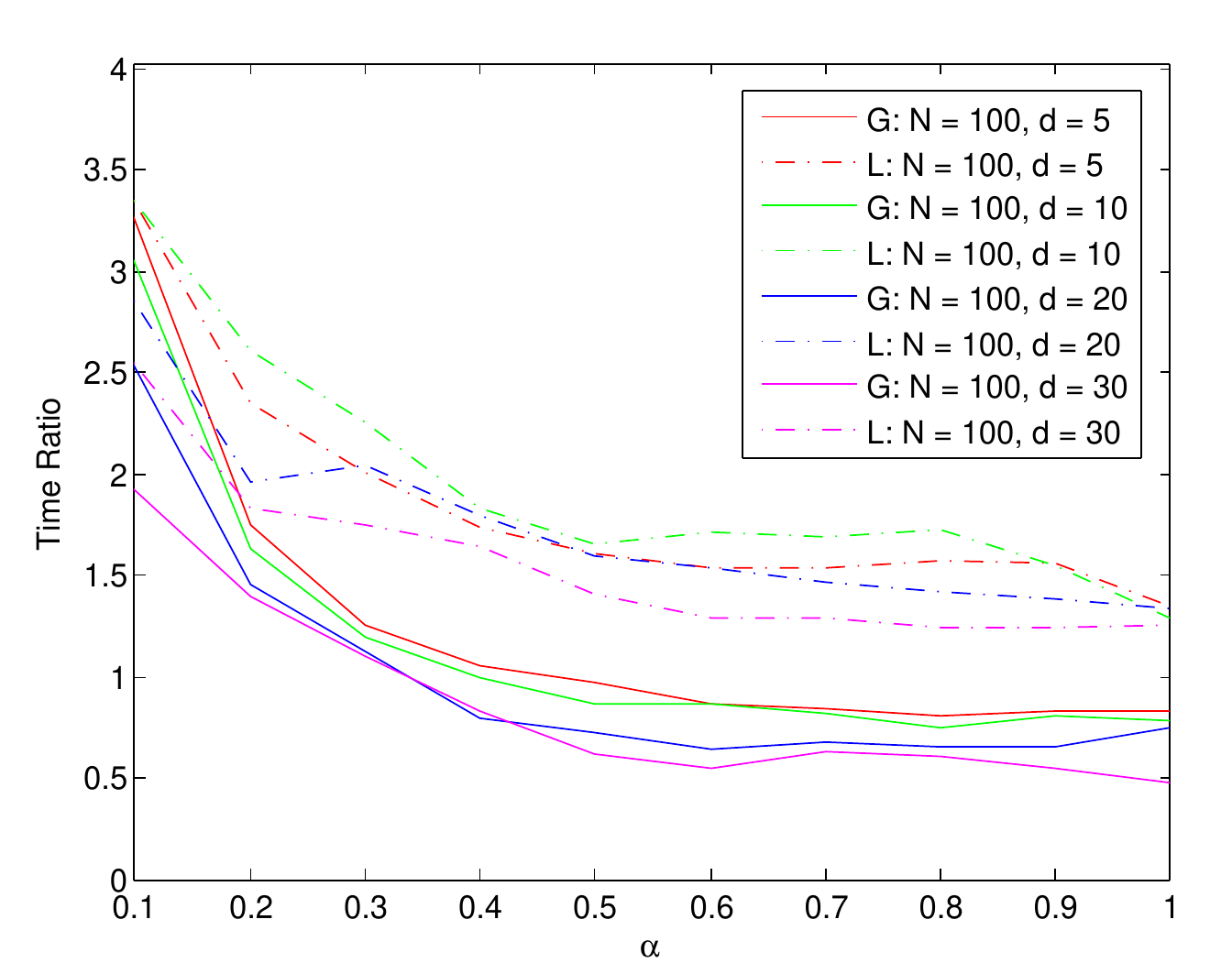}}
    \caption{(a) shows an example of a uniform-degree network. 
      (b) and (c) show the simulation performance
      obtained on uniform-degree networks with varying values of 
      $\alpha$, respectively in terms of cost ratios and
      time ratios. 
		Here, solid and dash-dot lines represent the results obtained by the selective communication scheme using global optimization and local optimization 	
		respectively.
		The different curves represent the results with different 	
		degrees.}
    \label{fig:Unet}
\end{figure*}

\begin{figure*}
    \centering
    \subfigure[Topology of an example network]
    {\includegraphics[width=0.31\textwidth]{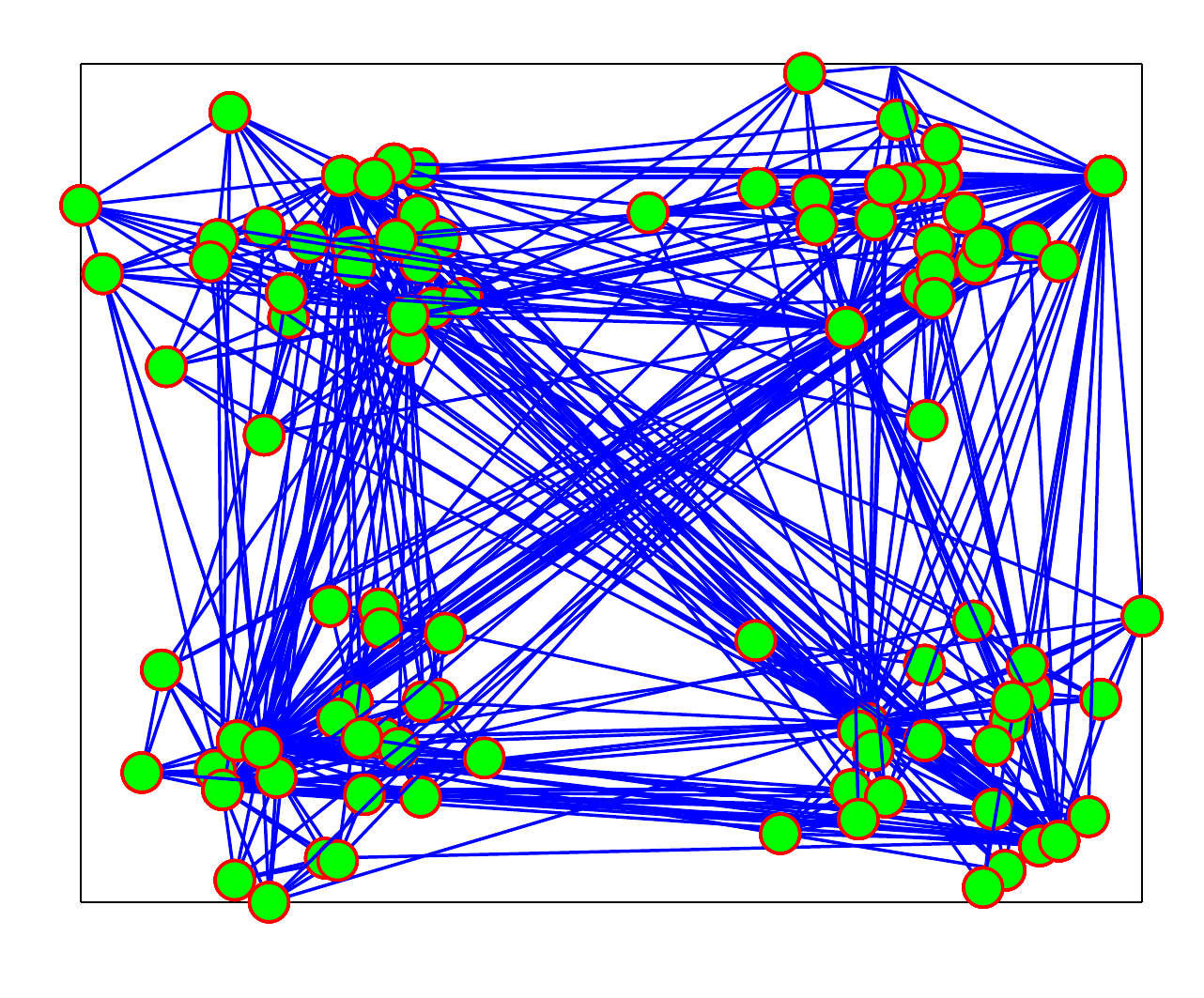}}
    \subfigure[Cost ratio as a function of $\alpha$]
    {\includegraphics[width=0.31\textwidth]{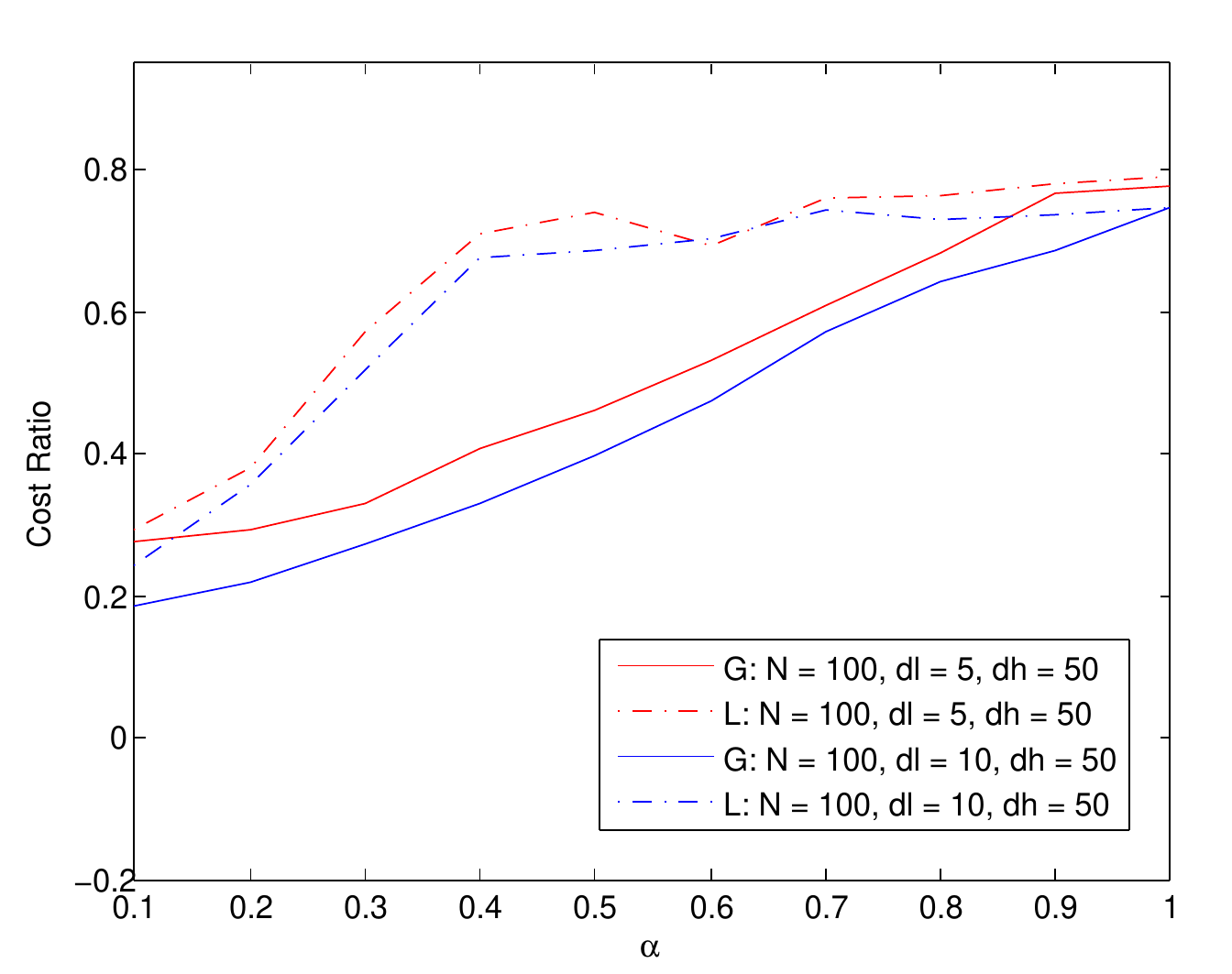}}
    \subfigure[Time ratio as a function of $\alpha$]
    {\includegraphics[width=0.31\textwidth]{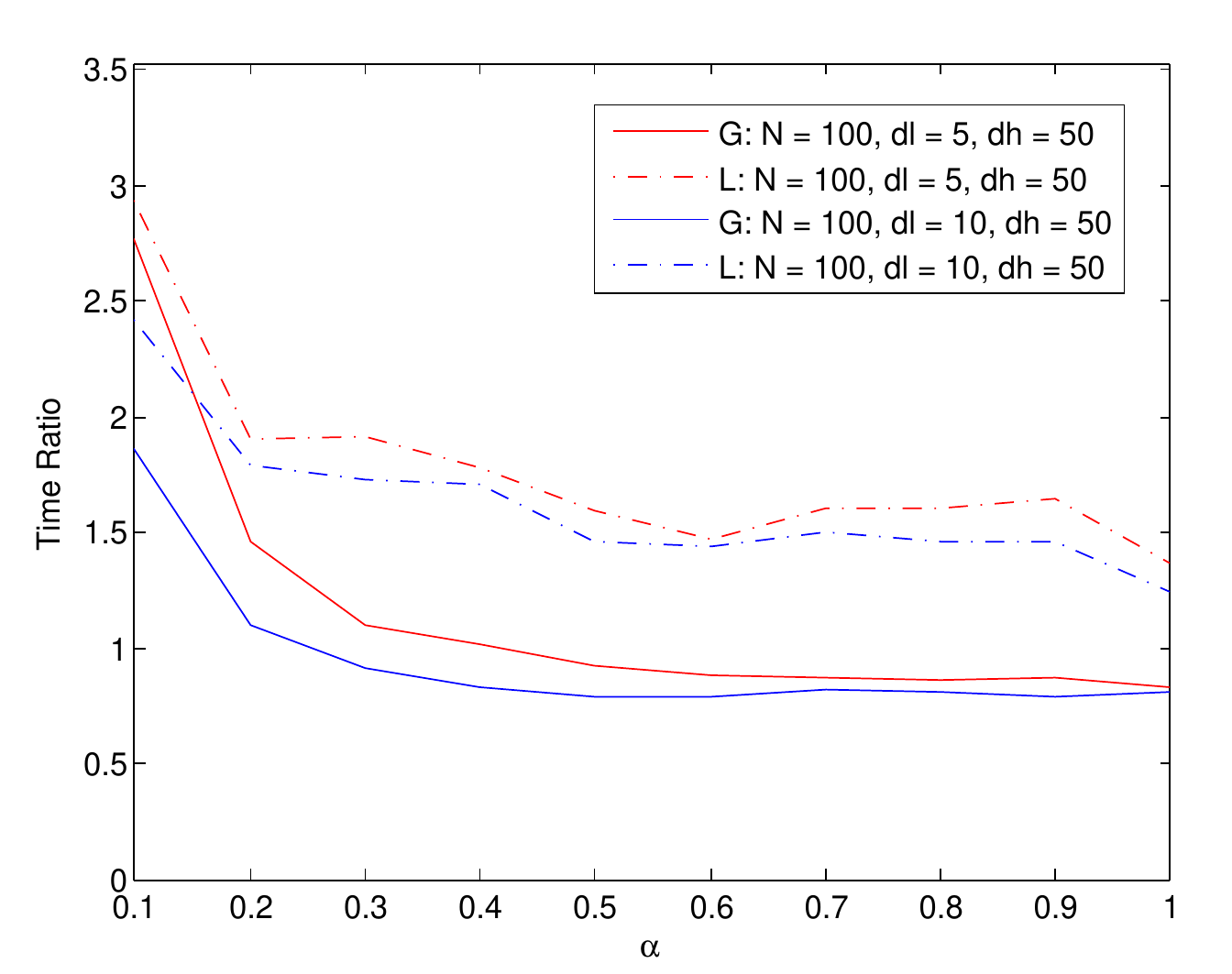}}
    \caption{(a) shows an example of a nonuniform-degree network. 
      (b) and (c) show the simulation performance
      obtained on uniform-degree networks with varying values of 
      $\alpha$, respectively in terms of cost ratios and
      time ratios.
 		Here, solid and dash-dot lines represent the results obtained by the selective communication scheme using global optimization and local optimization 	
		respectively.
		The different curves represent the results with different 	
		degrees.}
    \label{fig:Snet}
\end{figure*}

To test the practical performance of the proposed approach, 
we conducted simulations on synthetic networks for
both the globally and locally optimal link selection methods 
(respectively presented in Sections~\ref{sec:globalopt} and \ref{sec:localopt}).

In the simulation, we consider four types of networks: (1) \emph{uniform-degree networks}, where all nodes have similar
degrees, \ie~the number of neighbors; (2) \emph{nonuniform-degree networks}, where a small portion of nodes have substantially higher degrees than others; (3) \emph{star networks}, where a central node is connected to all the other nodes; and (4) \emph{chain networks}, where all the nodes are connected in a pattern of chain. 

For each network type, we randomly produced the network topology and conducted simulations under a variety of configurations. The step size, $\delta$, was found using Eq. (\ref{eq:optstep}), which achieves the fastest asymptotic convergence rate when all the links are used in the iterative updates. The upper bound in the cost constraint (Eqs. (\ref{eq:global_cc})
and (\ref{eq:lc})) were set to different values in our test, so as
to study the trade-off between energy cost and convergence rate
with the proposed schemes.  

The simulations were conducted as follows. 
First, an initial state value $x_v(0)$ was independently drawn from a
standard normal distribution ($\sigma^2 = 1$) for each node of the
network.
With the initialized states, we ran the iterative updating procedure, 
using the adaptive scheme to choose a subset of links at each
iteration.
The updating procedure was stopped when the difference between the
highest and lowest state values were below a tolerance value
$\varepsilon$, as
\begin{equation} \label{eq:tolv}
    \|\max(\vx(t)) - \min(\vx(t))\| < \varepsilon
\end{equation}
For this work, the tolerance $\varepsilon$ was set to $10^{-3}$.
Along with the simulation of an adaptive scheme, we also
ran a simulation upon the same network and same initial states without
adaptation (using all links at every iteration) to establish
the baseline for comparison. 
Let $m$ be the number of edges. Then for
global optimization, the value $C$ in the cost constraint (see
Eq. (\ref{eq:global_cc})) is set to $\alpha m$. The critical parameter in our work is $\alpha$ which is the ratio of the maximum number of links allowed at each iteration to the total number of available links. Similarly, the value $C_v$ for local optimization (see Eq. (\ref{eq:lc}) is set to $\alpha d_v$, where $d_v$ is the degree of node $v$. 

We consider two different metrics to assess the performance of the selective communication scheme: 
(1) \emph{Cost}: the total communication cost, i.e., the sum of the costs of all iteration (the cost of each iteration is defined to be the number of edges used in that iteration);
(2) \emph{Time}: the number of iterations that is needed to attain consensus.
We also evaluate the ratios of the time and cost obtained using the selective schemes to that obtained by the baseline using all available links at each iteration, in order to quantitatively measure how link selection influences the performance. The cost ratio measures how much the selective scheme improves the energy efficiency under the given setting, while the time ratio reflects how much the use of the selective scheme influences the convergence rate.

\subsection{Simulation on Uniform-Degree Networks}
\label{sub:uniform}
We first studied the proposed algorithms on uniform-degree networks.
As illustrated in Fig~\ref{fig:Unet}(a), a uniform-degree network used in this simulation contains $n = 100$ nodes, each connecting to five nodes on average.

We tested both globally optimal and locally optimal selection schemes on
uniform-degree networks. Fig~\ref{fig:Unet}(b) and Fig~\ref{fig:Unet}(c) respectively shows the
cost ratios and the time ratios obtained with different $\alpha$
values, and different degrees $d$.
From the results, we observe the following. First, both schemes can achieve substantial improvement on energy efficiency, with the convergence time maintained at a reasonable level. Second, the cost ratio decreases and the time ratio increases, as the value of $\alpha$ decreases. For example, under the setting with $d = 20$, when $\alpha$ is reduced from $0.8$ to $0.3$, the cost ratio obtained with global optimization decreases from $0.4$ to $0.3$, while the time ratio increases from below $1.0$ to $1.2$. With local optimization scheme, the cost ratio decreases from $0.7$ to $0.45$, while the time ratio increases from $1.5$ to $2.0$. Third, greater improvement is achieved on the network with higher degrees, i.e., each node has more neighbors. Intuitively, in such networks, edges can be more redundant. As a consequence, we can suppress more edges at each iteration, without noticeably affecting the convergence.
 
\subsection{Simulation on Nonuniform-Degree Networks}
\label{sub:nonuniform}

Next, we performed simulations on another type of networks,
the non-uniform degree networks, to see whether the proposed methods
exhibit different behaviors. 

As shown in Fig~\ref{fig:Snet}, a non-uniform network is composed of 
four clusters of nodes, each with $25$ nodes. Each cluster has two
high-degree nodes that link to $50$ other nodes, among which $24$ are 
within the same cluster, and $26$ are in others. Other nodes have much lower degree ($5$ on average) and only connect to nodes in the same cluster. 
Such networks can be considered as simplified versions of the hierarchical networks, where the high-degree nodes play a crucial
role for both within-cluster consensus and cross-cluster state
propagation. Thus the selection of the links that are incident
with such nodes is important.

Fig~\ref{fig:Snet}(b) and Fig~\ref{fig:Snet}(c) respectively show the
cost ratios and time ratios, obtained with different $\alpha$ values. 
Again, both adaptive schemes (global optimization and local
optimization) yield significant improvement on energy efficiency. For
example, when $\alpha = 0.3$, the scheme using global optimization
saves about $70\%$ of the communication cost, with slightly increased
convergence time  (the time ratio is $1.0 - 1.2$). Under the same
setting, the scheme with local optimization can save about $60\%$ of
the energy, while the convergence time is still maintained at a
reasonable level (time ratio is about $1.7$).
In addition, we observe similar trends from the results, 
\eg~the cost ratio decreases and the time ratio increases as $\alpha$
increases.

\subsection{Performance with Link Failures}
\label{sub:link_failure}

\begin{figure}[t]
    \centering
    \subfigure[Cost as a function of link fail probability $p$]
    {\includegraphics[width=0.31\textwidth]{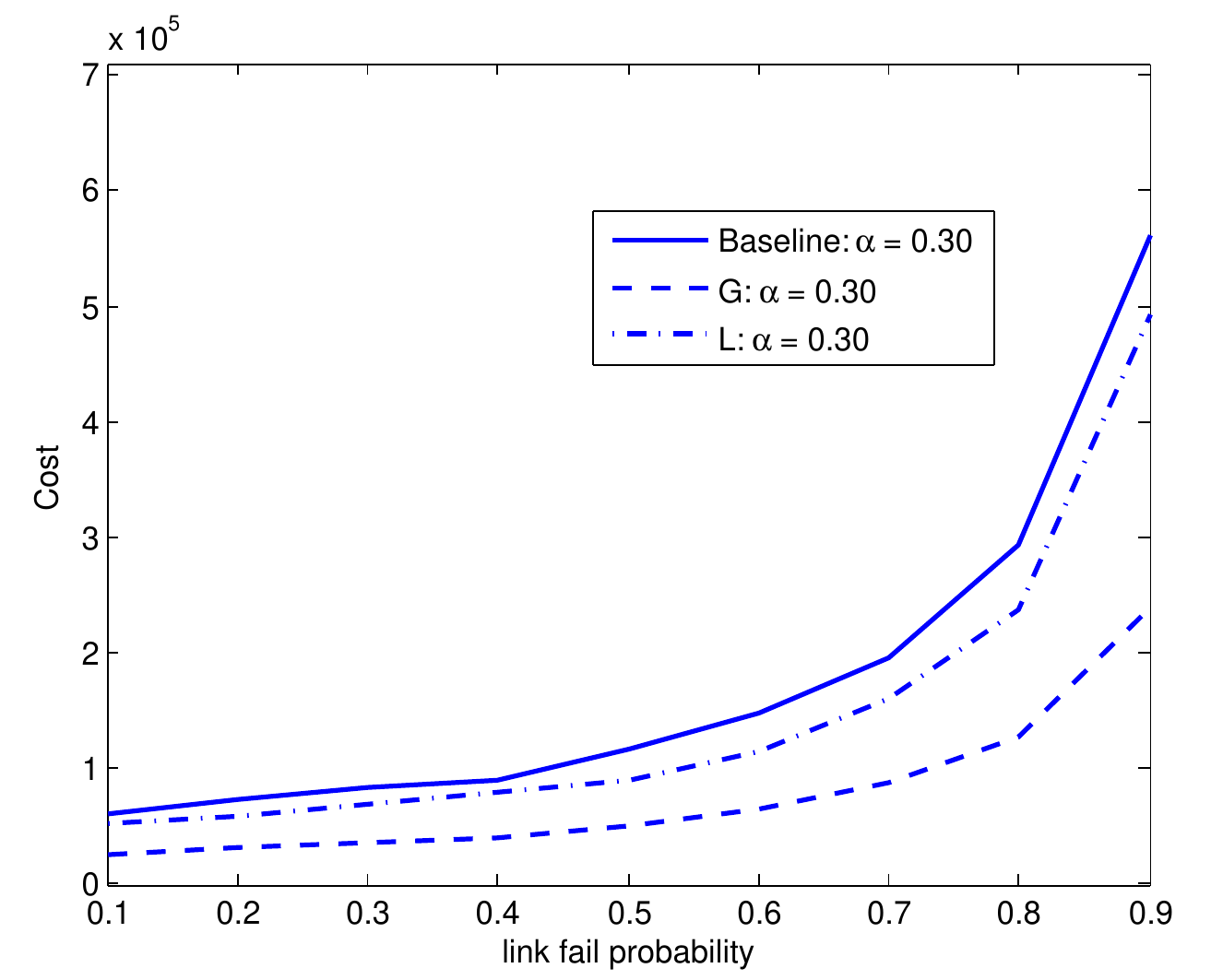}}
    \subfigure[Time as a function of link fail probability $p$]
    {\includegraphics[width=0.31\textwidth]{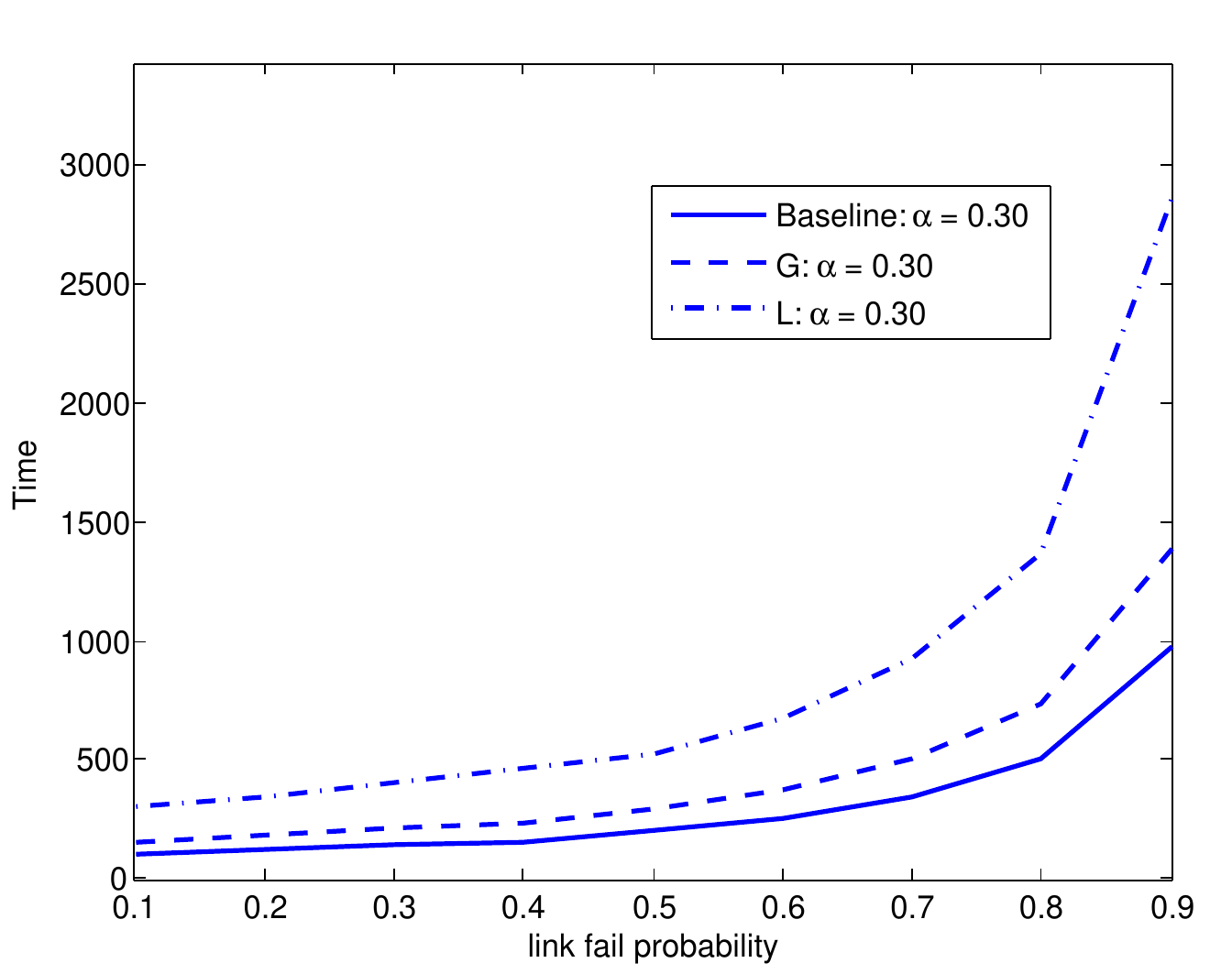}}
    \caption{The performances obtained with varying link failure probability. 
		Here, solid, dashed, and dash-dot lines respectively represent the results 	
		obtained using baseline scheme (without link selection), global 	
		optimization, and local optimization.}    
    \label{fig:probability}	
\end{figure}

We also investigate how the algorithms work in the presence of link failures. Sensor networks, especially those deployed in field, are often subject to communication failures, due to various causes, such as adversarial environmental changes, channel interference, etc. Therefore, the practical effectiveness of a communication scheme is influenced, to a great extend, by its robustness against such problems.

We study this issue via a simulation under a simplified setting, where we assume that each communication link fails independently with a certain probability, we randomly turned off some links at each communication step. We ran the simulations on a uniform degree network of $d = 10$ similar to Fig~\ref{fig:Unet}(a) and $\alpha$ is set to $0.3$, with different probabilities of failure, ranging from $0.1$ to $0.9$.

To take a more close examination of how the link failure affects performance, we plot the absolute cost and convergence time for both schemes, as shown in Fig~\ref{fig:probability}, instead of their ratios. 
The results clearly show as expected that for all schemes, both communication cost and convergence time increases as the probability of link failure increases. 
However, regardless of the variation of failing probabilities, the proposed link selection schemes consistently achieve notable improvement on energy efficiency. 
Particularly, the scheme using global optimization can save over $60\%$ of the total cost, while the one using local optimization can save over $20\%$. 
In addition, we observe that while the convergence time increases as the communication links become more likely to fail, it is still maintained at a reasonable level, as compared to the baseline.  
With these results, we contend the selective communication schemes are robust to link failures.

\subsection{Simulation on Star and Chain Networks}
\label{sub:star}

\begin{figure*}
    \centering
    \subfigure[Topology of an example star network]
    {\includegraphics[width=0.31\textwidth]{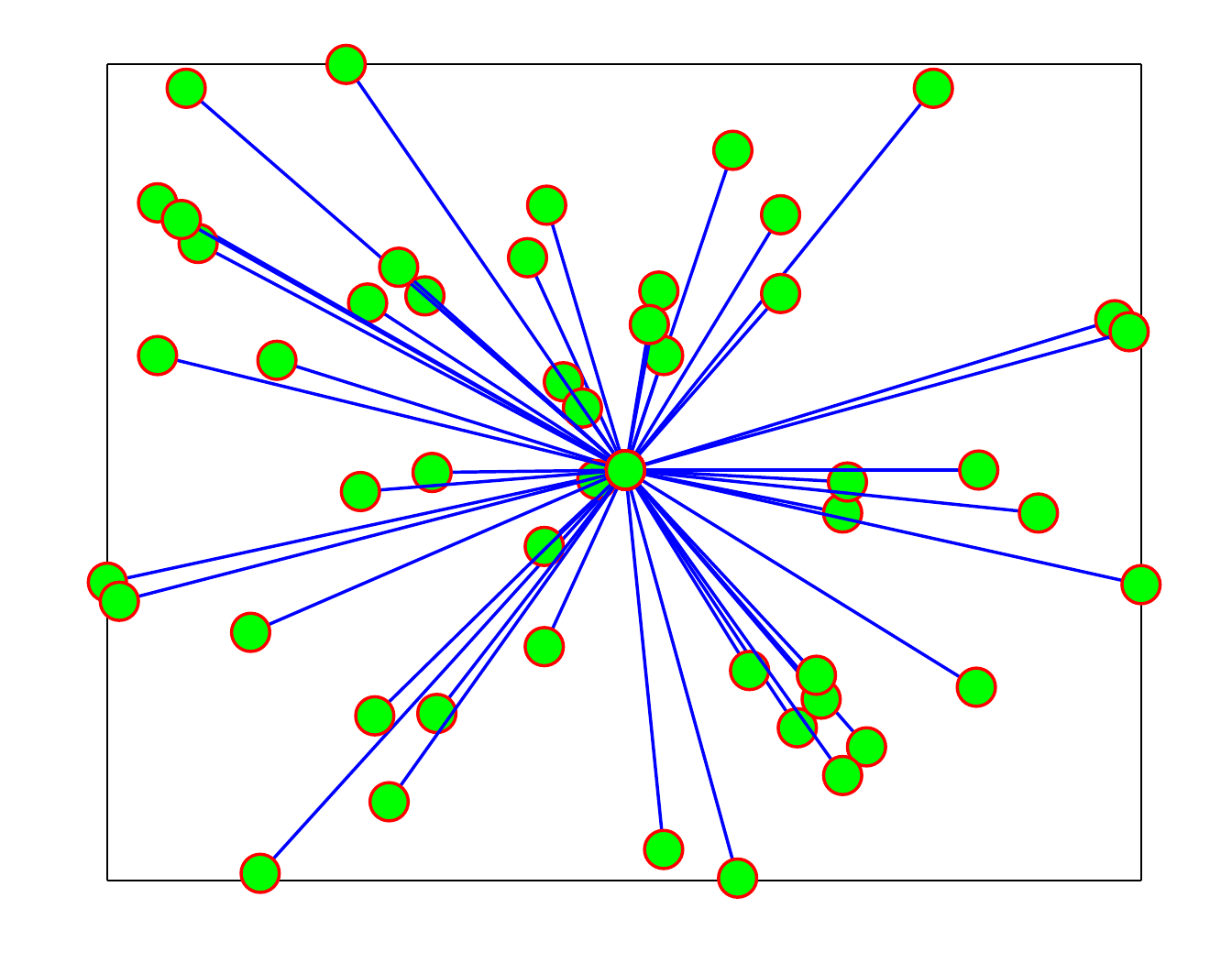}}
    \subfigure[Cost as a function of the number of nodes $n$]
    {\includegraphics[width=0.31\textwidth]{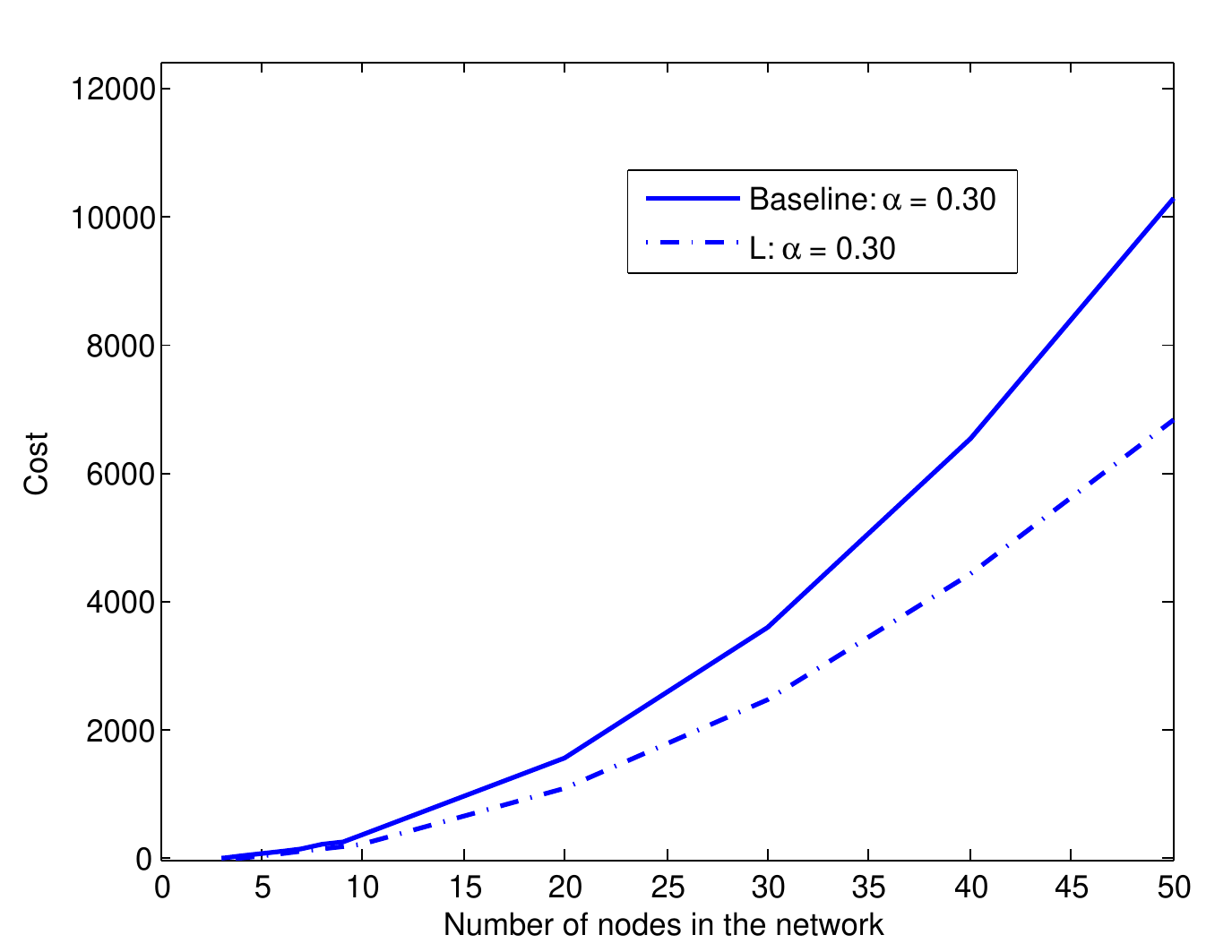}}
    \subfigure[Time as a function of the number of nodes $n$]
    {\includegraphics[width=0.31\textwidth]{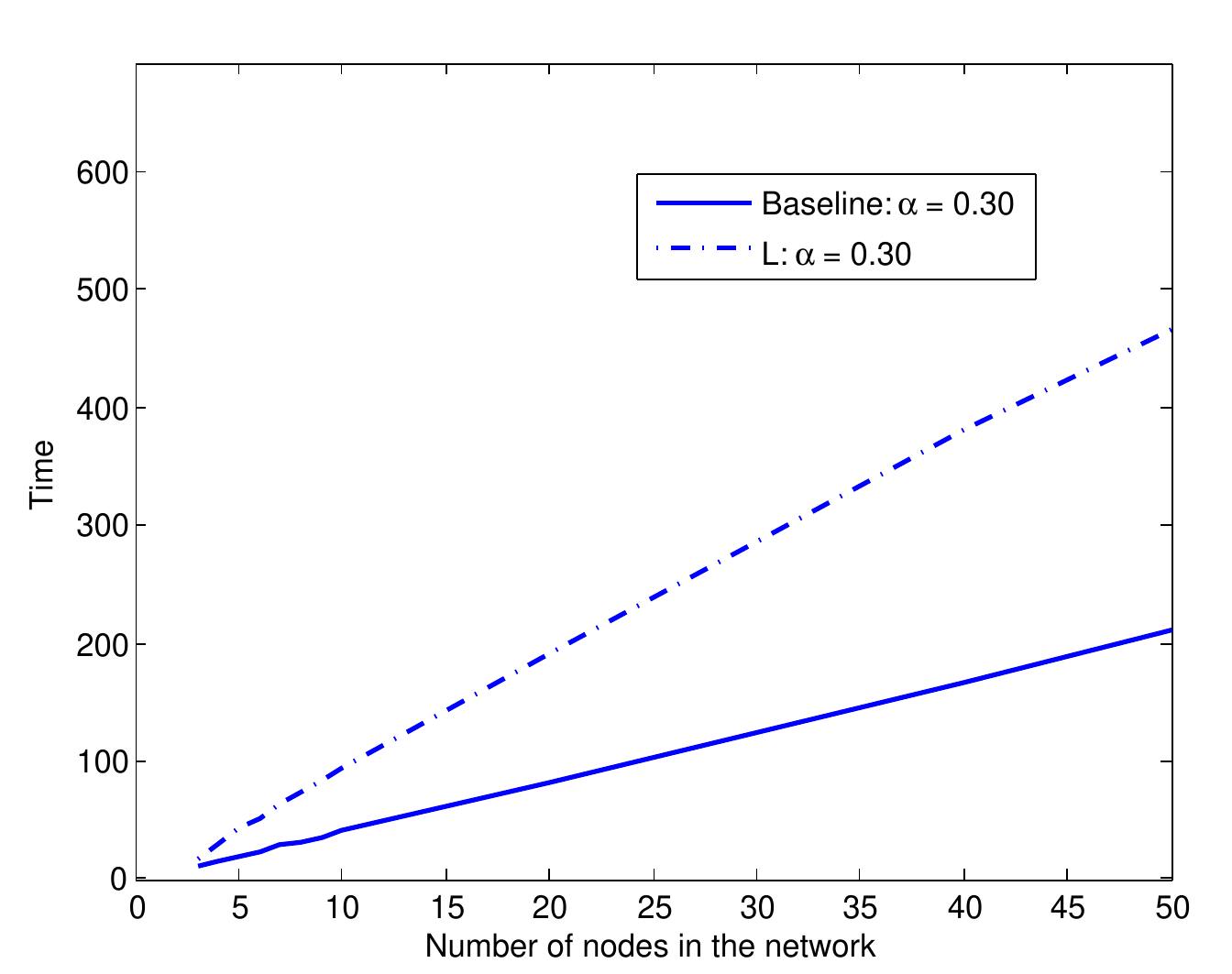}}
    \caption{(a) shows an example of a star network. 
      (b) and (c) show the simulation performance
      obtained on star networks with varying number of nodes $n$, respectively 	
		in terms of cost and time.
		Here, solid and dash-dot lines 	
		respectively represent the results obtained using baseline scheme
		(without link selection) and local optimization.}
    \label{fig:star}
\end{figure*}

\begin{figure*}
    \centering
    \subfigure[Topology of an example chain network]
    {\includegraphics[width=0.31\textwidth]{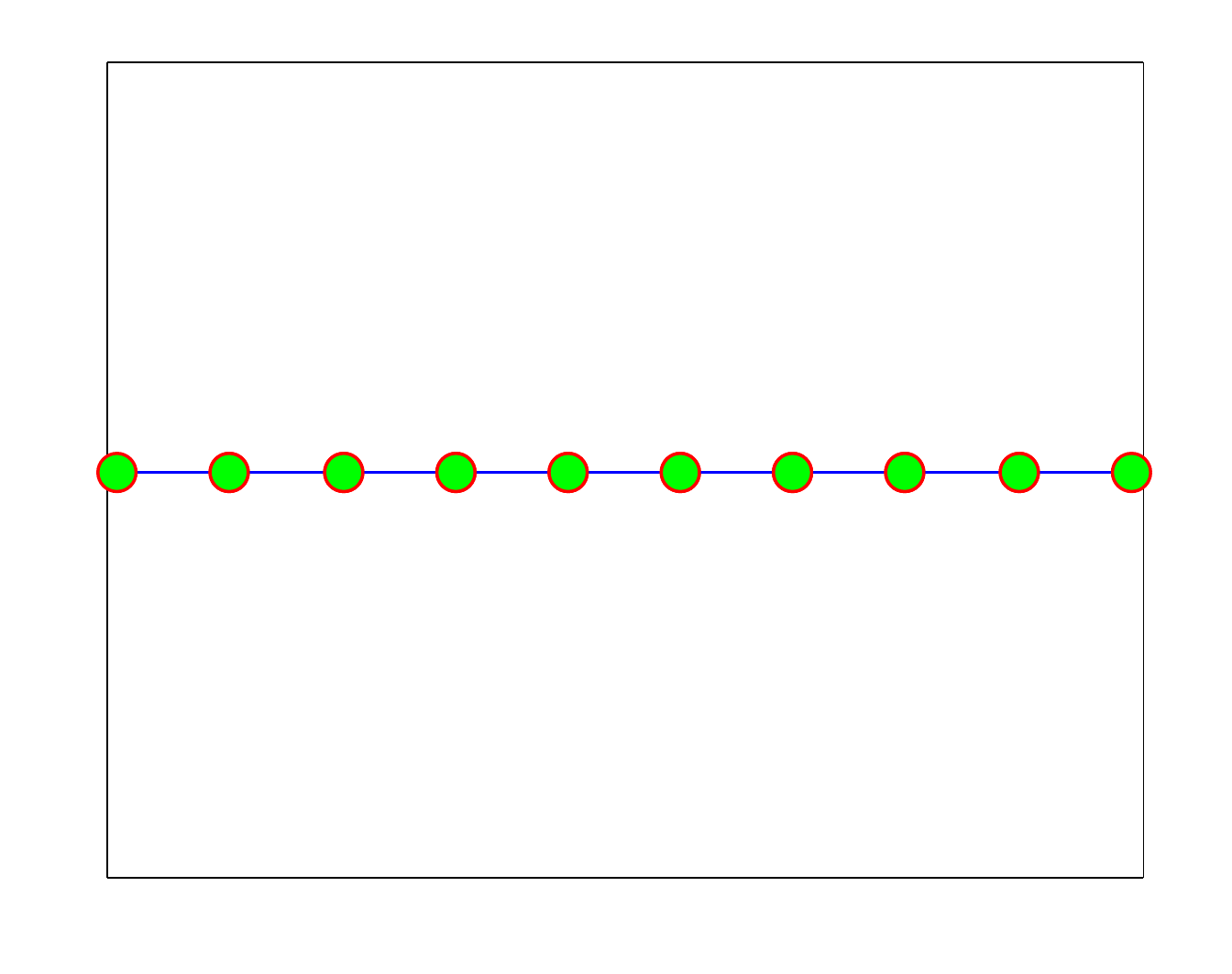}}
    \subfigure[Cost as a function of the number of nodes $n$]
    {\includegraphics[width=0.31\textwidth]{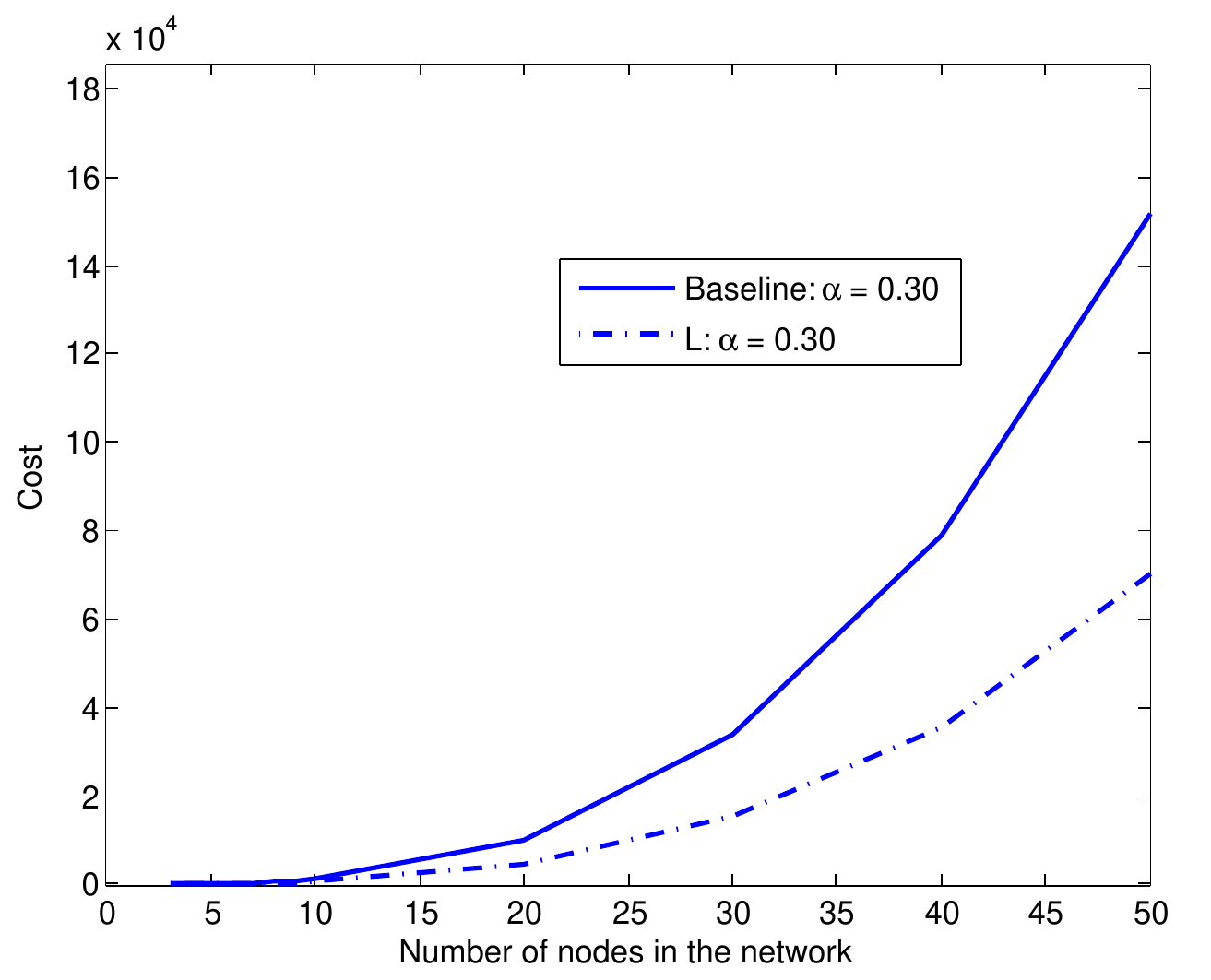}}
    \subfigure[Time as a function of the number of nodes $n$]
    {\includegraphics[width=0.31\textwidth]{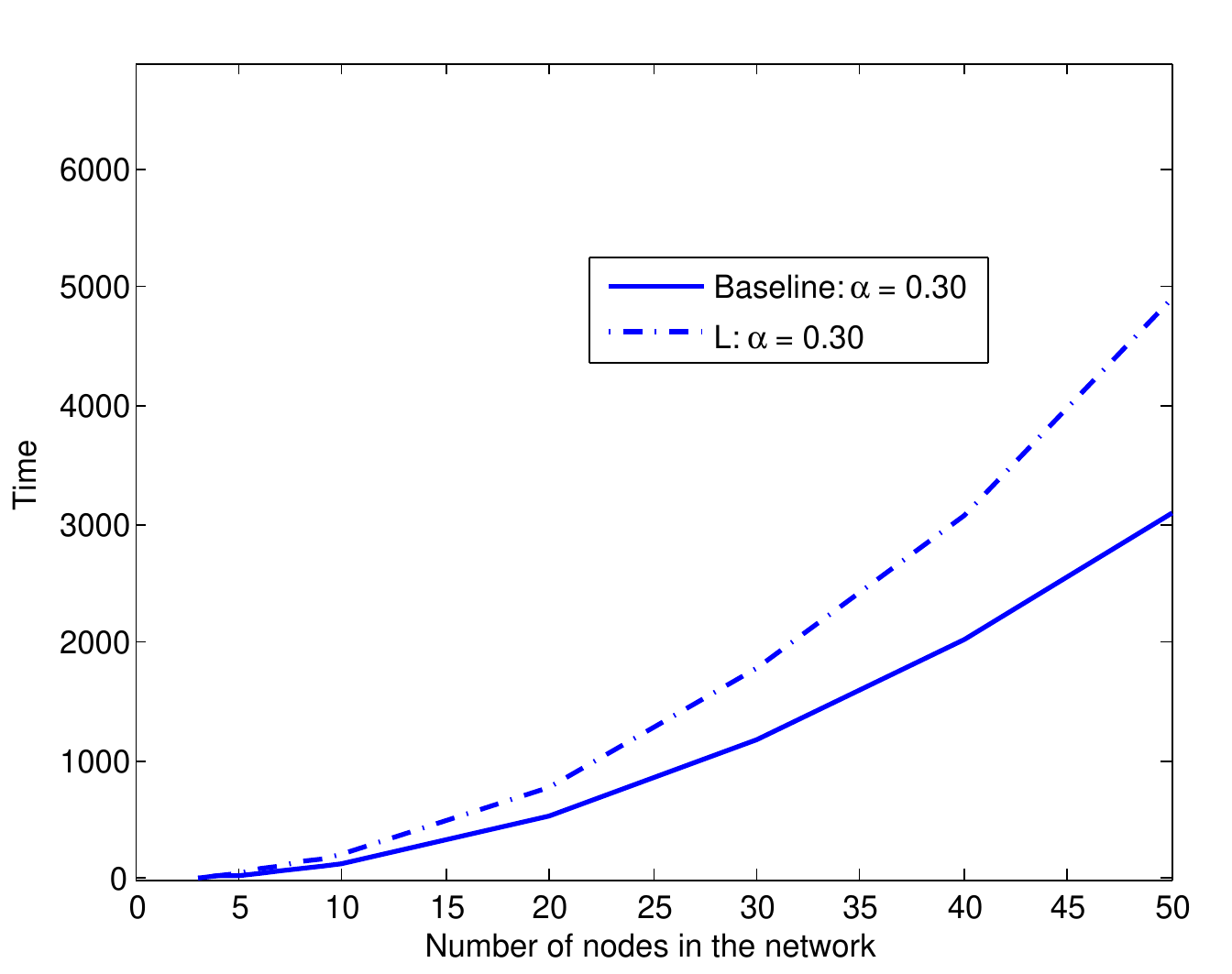}}
    \caption{(a) shows an example of a chain network. 
      (b) and (c) show the simulation performance
      obtained on chain networks with varying number of nodes $n$, respectively 	
		in terms of cost and time. 
		Here, solid and dash-dot lines respectively represent the results 	
		obtained using baseline scheme (without link selection) and local optimization.}
    \label{fig:chain}
\end{figure*}

Finally, we present a more detailed study on two special types of networks: star networks and chain networks. A star network consists of a central node, which is directly connected to all other nodes. A chain network consists of nodes arranged in form of a chain, with each node connected to at most two nodes. These two types of networks represent two topological extremes. In a star network, information is quickly exchanged between center and leaf nodes, while in a chain network, information is propagated much more slowly, as it has to pass along the chain, from one end to the other.
Fig~\ref{fig:star}(a) and Fig~\ref{fig:chain}(a) respectively illustrate a star network with $50$ nodes and a chain network with $10$ nodes.

In this experiment, we considered star networks with different number of nodes, with an aim to study how network structure affects energy efficiency and network performance. Here, we focus on local scheme, as it is practically feasible. Empirically, we set $\alpha$ to $0.3$ in the simulation, which means up to $30\%$ of the total available links are allowed at each iteration.

The results obtained on star networks are shown in Fig~\ref{fig:star}. Specifically, Fig~\ref{fig:star}(b) and Fig~\ref{fig:star}(c) respectively show the total communication costs and the number of iterations needed to achieve convergence, both as functions of the network size $n$. Fig~\ref{fig:chain} shows the results on chain networks, in the same fashion.

The results of star and chain networks present similar trends. When the network size increases, the communication cost rises and it takes more time to reach consensus. More importantly, we also observe significant reduction of the total communication cost, with the use of local optimization scheme. Note that it takes longer to achieve convergence as compared to the baseline setting. This is not surprising, as only a small subset of links are activated at each communication step. In particular, when the network size is 50, about $35\%$ and $55\%$ of cost can be saved for star network and chain network respectively, and it takes $2.2$ and $1.6$ of the convergence time accordingly. It is worth emphasizing that the value of this work is that it provides an effective way for one to trade off convergence performance for energy efficiency. This is very useful under many practical circumstances, where efficient use of energy may be more important than achieving the optimal convergence time.


\section{Conclusion}
\label{sec:conclusion}

In this paper, we presented two approaches to improve the
energy efficiency of distributed average consensus. Specifically, we
revisited the analysis of the consensus process, and established it as
a distributed optimization procedure that minimizes the value of
$\vx(t)^T \mL \vx(t)$. Motivated by this relation, we first developed
an approach that chooses an optimal subset of links for
communication at each iteration, by minimizing this value subject to a
cost constraint. Then, via approximation, we factorized this problem
into a set of sub-problems that can be solved locally, resulting in an approach that is practical in a distributed context.
In solving this problems, we derive a simple yet effective method,
which first solve a relaxed QP problem, and then obtain a subset of
links by random sampling.

We performed simulations on various types of networks to test the performance of the proposed schemes, and
compared results with a baseline scheme without link selection. The results acquired on uniform, non-uniform, star and chain networks clearly demonstrate that the proposed methods have consistently improved energy efficiency for the distributed consensus problem. Based on these results, we contend that dynamic link selection for distributed computing in sensor networks should be considered as a viable methodology to improve the energy efficiency of sensor networks.
We also studied the influence of the link reliability on the performance of the proposed methods. Our results provide practical guidance as to how to choose design parameters in order to strike a balance between convergence speed and energy cost.



%

\appendices


\ifCLASSOPTIONcaptionsoff
  \newpage
\fi


\bibliographystyle{IEEEtran}
\bibliography{reference2013}

\end{document}